\documentclass[twocolumn,amsmath,amssymb,showpacs,superscriptaddress]{revtex4}%
\usepackage{epsf}
\usepackage{epsfig}
\usepackage{graphicx}
\usepackage{dcolumn}
\usepackage{bm}
\usepackage{amsmath}
\usepackage{amsfonts}
\usepackage{amssymb}

\usepackage[usenames,dvipsnames]{color}

\newcommand{\unit}[1]{\ensuremath{\, \mathrm{#1}}}

\providecommand{\U}[1]{\protect\rule{.1in}{.1in}}
\input epsf
\begin{document}
\title{Enhancing single-parameter quantum charge pumping in carbon-based devices}
\author{Luis E. F. Foa Torres}
\affiliation{Instituto de F\'{\i}sica Enrique Gaviola (IFEG-CONICET) and
FaMAF, Universidad Nacional de C\'{o}rdoba, Ciudad Universitaria, 5000
C\'{o}rdoba, Argentina.}
\author{Hern\'{a}n L. Calvo}
\affiliation{Instituto de F\'{\i}sica Enrique Gaviola (IFEG-CONICET) and
FaMAF, Universidad Nacional de C\'{o}rdoba, Ciudad Universitaria, 5000
C\'{o}rdoba, Argentina.}
\affiliation{Institut f\"{u}r Theorie der Statistischen Physik, RWTH Aachen University, D-52056 Aachen, Germany.}
\author{Claudia G. Rocha}
\affiliation{Department of Physics, NanoScience Center, University
of Jyv\"askyl\"a, 40014 Jyv\"askyl\"a, Finland}
\affiliation{Institute for Materials Science and Max Bergmann Center of Biomaterials,
Dresden University of Technology, D-01062 Dresden, Germany}
\author{Gianaurelio Cuniberti}
\affiliation{Institute for Materials Science and Max Bergmann Center of Biomaterials,
Dresden University of Technology, D-01062 Dresden, Germany}
\affiliation{Division of IT Convergence Engineering, POSTECH, Pohang 790-784, Republic of Korea}
\date{\today}

\begin{abstract}
We present a theoretical study of quantum charge pumping with a
single ac gate applied to graphene nanoribbons and carbon nanotubes
operating with low resistance contacts. By combining Floquet theory with Green's function formalism, we show that the pumped current can be tuned and enhanced by up to two orders of magnitude by an appropriate choice of device length, gate voltage intensity and driving frequency and amplitude. These results offer a promising alternative for enhancing the pumped currents in these carbon-based devices.
\end{abstract}

\maketitle

%\title{Tunneling through driven carbon-based Fabry-P\'{e}rot devices: Manifestation
%of quantum wagon-wheel effect}

%\title{Tunneling through driven carbon-based Fabry-P\'{e}rot devices: Manifestation
%of quantum wagon-wheel effect}

A current flow in a two terminal conductor normally requires a
bias voltage difference between the electrodes.
However, in systems where the coherence length of the electrons
exceeds the device length, the use of ac fields can produce a
non-vanishing current even at zero bias voltage. This quantum
coherent effect is called \textit{quantum charge pumping}
\cite{Thouless1983,Altshuler1999,Buttiker2006} and has been observed
experimentally \cite{Switkes1999,DiCarlo2003,Blumenthal2007}. When
the driving frequency ($\Omega$) is low enough in such a way that
$\Omega \ll \tau^{-1}$ (being $\tau$ the time spent by an electron
to traverse the sample) then the transport is said to be adiabatic.
Under such conditions at least two out of phase time-dependent
parameters are necessary to achieve pumping \cite{Brouwer1998}.
Beyond the realm of the adiabatic approximation, pumping with a
single parameter is also possible
\cite{Vavilov2001,Moskalets2002,Arrachea2005,FoaTorres2005, Agarwal2007} as
confirmed by recent experiments
\cite{Kaestner2008,Fujiwara2008,Kaestner2009}. While a drawback for
such monoparametric pumps is that higher frequencies are required to
achieve similar currents, a big advantage comes from the reduction
in the number of necessary contacts which makes it promising for
scalable, low-dissipation, devices. In this way, single-parameter
pumping has become a quest for achieving higher currents in nanoscale
transmission channels.

The advent of graphene \cite{CastroNeto2009} and their lower
dimensional cousins, graphene nanoribbons and carbon nanotubes
\cite{Charlier2007}, provided an outstanding ground for exploring
the physics of quantum charge pumping as evidenced by recent
studies. Most of them are focused in
flat graphene samples driven in the adiabatic regime
\cite{Prada2009,RuiZhu2009,Blaauboer2010} or beyond the adiabatic regime but in the 
energy range close to the Dirac point \cite{SanJose2011,Gu2009}. The
observation of other prominent features such as van Hove singularities
usually requires an important gate voltage but recent experiments
performed on twisted graphene layers demonstrate that singularities can
be brought arbitrarily close to the Fermi energy \cite{GuohongLi2010},
opening interesting prospects for beyond-the-Dirac-point-studies.

In this Letter, we explore the interplay between electronic
structure and non-adiabatic effects in carbon-based quantum pumping
devices with a \textit{single} parameter. The transmission channels are
composed of carbon nanotubes or graphene nanoribbons but our
results are expected to be valid for generic quasi one-dimensional
systems. By modeling a device with low resistance contacts and
driven by a single time-dependent gate, we show that the pumped
current can be enhanced by up to two orders of magnitude by gating
the system close to a van Hove singularity. Moreover, it is shown that 
substantial improvement is also possible at the Dirac point when tuning the interplay
between driving frequency and device length. Our study points out
alternative directions to bring these devices closer to reality.

%Furthermore, the scaling with frequency of the maximum pumped current shows to
% be better than the typical quadratic one found for slowly varying
%%density of states. Our study points out new directions to bring these
%devices closer to reality.

\begin{figure}[ptbh]
\includegraphics[width=8.0cm]{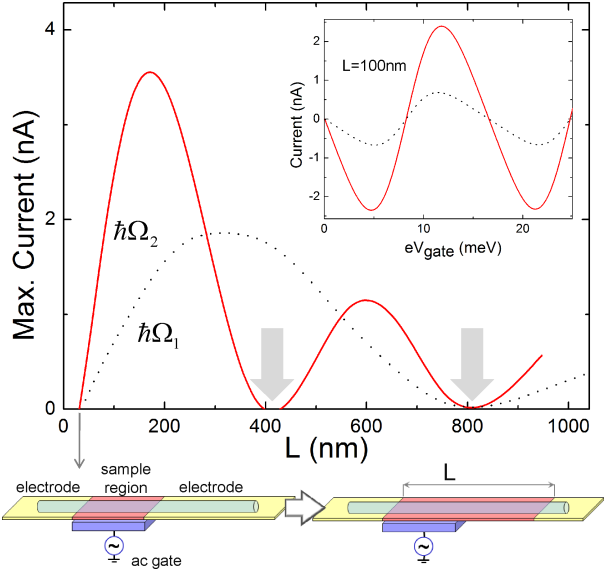} \caption{(color online)
Inset: Pumped current as a function of the gate voltage for a
$(24,0)$ carbon nanotube with $L=100\unit{nm}$. Main frame: maximum current
pumping as a function of the system length at two fixed driven
 frequencies: (solid red line) $\hbar \Omega_2 = 4.050\unit{meV}$ and
 (dotted line) $\hbar \Omega_1 = 2.025\unit{meV}$. The ac field intensity is
 set to $eV_{\mathrm{ac}}=1.35\unit{meV}$. The vertical down arrows (grey) mark the points where the pumped current vanishes. Bottom panels: schematic
 representation of the system with symmetric (left) and asymmetric (right) ac gate.}%
\label{fig1}%
\end{figure}

\textit{Tight-binding model and Floquet solution.} We consider a
device consisting of a graphene nanoribbon with passivated edges, or a single wall carbon nanotube, of
length $L$ connected to two semi-infinite electrodes. For the sake
of simplicity, the electrodes are considered to be a prolongation of
the sample located at the centre. The system is described through a
single $\pi$-orbital Hamiltonian \cite{Charlier2007} $H_{e}=\sum_{i}E_{i}^{{}}c_{i}%
^{+}c_{i}^{{}}-\sum_{\left\langle i,j\right\rangle }[\gamma_{i,j}c_{i}%
^{+}c_{j}^{{}}+\mathrm{H.c.}]$, where $c_{i}^{+}$ and $c_{i}^{{}}$
are the creation and annihilation operators for electrons at site
$i$, $E_{i}$ are the on site energies and $\gamma_{i,j}$ are
nearest-neighbors carbon-carbon hoppings. To simulate a system with
low resistance contacts, the central part of length $L$ (the
\textquotedblleft sample\textquotedblright) is connected to the
semi-infinite leads through matrix elements $\gamma_{t}$ smaller
than the carbon-carbon hopping \cite{Krompiewski2002}, $\gamma
=2.7\unit{eV}$ \ \cite{Charlier2007}. We set $\gamma _{t}=0.7\gamma$ in
order to mimic the Fabry-Perot conductance oscillations
\cite{footnoteFP} experimentally observed at low temperatures
\cite{Liang2001}.

In addition to the possibility of homogeneously gating the whole central
sample, which is useful for revealing Fabry-Perot oscillations as in
\cite{Liang2001}, here we include an ac gate asymmetrically
disposed on the scattering region as represented at the bottom of Fig. 1. The ac gate, which is
responsible for the dynamical breaking of the left-right symmetry,
is modelled through an additional on site energy term $E_{j\in
\mathrm{G}}=eV_{\mathrm{ac}}\cos(\Omega t)+eV_{gate}$ where $G$ is the region
where the ac gate is applied, $V_{\mathrm{ac}}$ is the amplitude of the ac gate,
and $V_{gate}$ is the one of a static gate which is applied to the whole sample, $\Omega$ is the driven
frequency. The current is obtained by using
Floquet theory \cite{Moskalets2002,Camalet2003,Kohler2005}. The dc component
of the time-dependent current $I(t)$ can be computed as
\begin{equation}
\bar{I}=\tfrac{2e^{2}}{h}\sum_{n}\int\left[  T_{R\leftarrow L}%
^{(n)}(\varepsilon)f_{L}(\varepsilon)-T_{L\leftarrow
R}^{(n)}(\varepsilon )f_{R}(\varepsilon)\right]  d\varepsilon,
\label{current}
\end{equation}
where $T_{R\leftarrow L}^{(n)}(\varepsilon)$ are the
transmission probabilities of the carriers coming from the left (L)
to right (R) electrodes which might absorb or emit $\mid n\mid$
photons depending if $n>0$ or $n<0$, respectively. These
transmission probabilities can be fully written in terms of the
Floquet Green's functions for the system as explained in \cite{FoaTorres2005}. The
number of Floquet channels ($n$) used in the calculation is chosen
by requiring the convergence of the transmission functions with at
least six significant digits. The electron distribution in the left
(right) lead is given by the Fermi function $f_{L}(\varepsilon)$
($f_{R}(\varepsilon)$).
The non-interacting model is justified when screening by the
surrounding gate or a metallic substrate reduces electron-electron
interactions, ruling out effects beyond
our present scope \cite{Guigou2007}. At the same time,
electron-phonon interactions can be ruled out at low bias voltages
($<150\unit{meV}$) since the relevant inelastic mean free path for
scattering with acoustical phonons is on the order of 1-2
micrometers at $300\unit{K}$ \cite{Roche2007}. Recently, a similar
method was applied to unveil laser-induced band gaps in graphene and their effect on
charge transport \cite{Calvo2011}. 

\textit{Single-parameter pumping: interplay between electronic
structure and non-adiabatic effects.} Although in the following we
present results for metallic nanotubes, the same physics was found
for armchair edge graphene nanoribbons. This is consistent
with the argument proposed in \cite{Rocha2010} based on the use
of a mode decomposition scheme for the electronic states.
Furthermore, we expect the trends found here to be valid for a
generic quasi one-dimensional system. Figures 1 and 2 show the
maximum pumped current calculated for a $(24,0)$ carbon nanotube
with an asymmetrical ac gate (see bottom schemes in Fig. 1) extending over a typical length $L_{\mathrm{ac}}=30\unit{nm}$. 

When pumping close to the charge neutrality point (CNP), the
Fabry-Perot oscillations manifest as an oscillatory pumped current
as shown in the inset of Fig. 1. Its period is commensurate to the
energy level spacing, $\Delta$, which characterizes the system
spectrum. Changing the device length allows the tuning of the energy
level separation since $\Delta \propto 1/L$. On Fig. 1 (main frame),
one can notice that the maximum pumped current can be strongly
modulated by the device length. The solid red line corresponds
to a driving frequency which doubles the one for the dotted line. The
zeroes in these curves (marked with down arrows) can be understood in terms of a
\textit{wagon-wheel effect} \cite{FoaTorres2009,Rocha2010}: whenever
the driving frequency is commensurate with the level spacing, the
system behaves as in the static regime and no current is pumped
through the system. A similar phenomenon was noted in \cite{Moskalets2002} for a double barrier structure. Another
feature in the decrease of the successive maxima in Fig. 1 (main frame). 
To explain this issue, first one notes that the
pumped current is obtained by integrating a kernel function given in Eq. \ref{current}, which in our case turns out to be
 periodic as demonstrated on Fig. 1 inset. While the kernel amplitude remains
 approximately constant when changing the device length, the relevant energy scale over which the integration is performed
 is $min(\hbar\Omega, \Delta)$. This leads, for big enough $L$ to a series of maxima
which scale with $\Delta \propto 1/L$ as seen in Fig. 1 (main frame).

\begin{figure}[ptbh]
\includegraphics[width=8.0cm]{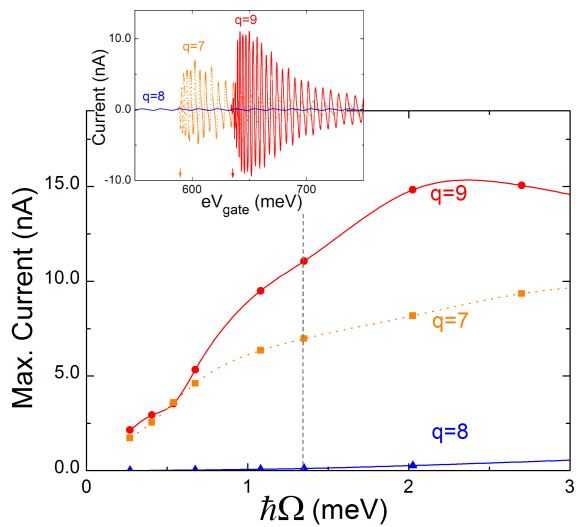} \caption{Maximum current pumping
 obtained for (24,0) carbon nanotube. (color online) Inset: Pumped
current as a function of the gate voltage. The different curves are
the contributions from the different sub-bands. The position of the
van Hove singularities are indicated with arrows on the horizontal
axis. The driving frequency used in the inset is marked with a vertical dashed line in the main frame. 
Main frame: Maximum pumped current for each contributing
sub-band as a function of the driving frequency. $eV_{\mathrm{ac}}$
 is set as in Fig. 1 and $L=100\unit{nm}$.}% 
\label{fig2}%
\end{figure}

Beyond the CNP, application of a gate voltage can unveil interesting
features when tuned nearby van Hove singularities (vHS). 
The physics in this case becomes more transparent if the total current is split into 
the contributions due to the different subbands. This is achieved by resorting 
to a unitary transformation which diagonalizes the electronic Hamiltonian for each layer perpendicular to the transport direction \cite{Mingo2001}.
For a $(N,0)$ nanotube, the different subbands correspond to linear chains with alternating hoppings $\gamma_{0}$
and $\gamma_{q}=2\gamma_{0}\cos(q\pi/N)$ ($q=0,1...,N-1$).  %with dispersion
%$relations: $\varepsilon^{(0)}(k)=\pm\sqrt{\gamma_{0}^{2}+\gamma_{q}%
%^{2}+2\gamma_{0}\gamma_{q}\cos(3ka_{cc}/2)}$. 
When $N$ is an integer multiple
of three, the metallic subbands correspond to $q=N/3,2N/3$.
Figure 2 shows the current pumped as a function of the gate voltage $V_{gate}$ (inset) and
driving frequency (main frame) applied to a (24,0) nanotube
for a typical set of parameters. Different curves show the
contributions coming from each subband with $q=7,8,9$ to the total
current (since the contributions for $q=15,16,17$ are equal to those shown here we concentrate only in the latter). 
The curve labeled as $q=8$ corresponds to one of the two linear bands
which crosses the CNP and which is responsible for the metallic
character of the tube. The remaining curves, $q=7$ and $q=9$, lead
to the first and second van Hove singularities, respectively. By
inspection of the inset one sees that subbands labeled with
$q=7,9$, which reveal vHSs within the displayed gating range,
dominate over the one with $q=8$. Furthermore, the contribution due to $q=7,9$ show a damping
behaviour, revealing a maximum intensity close to the corresponding
vHS and then a decay. What triggers such a dramatic increase of the
current shown in Fig. 2 inset? Careful analysis of the numerical results show that as the
gate voltage is tuned closer to the vHS, there is a crossover from a
regime where inelastic effects (which are responsible for
non-adiabatic pumping) are weak to one where they become dominant
(close to the vHS where the level spacing becomes smaller than the driving frequency
$\hbar\Omega$ ). 

Another relevant point is the scaling of the pumped current with the
driving frequency. In contrast to adiabatic pumping, where the
frequency can be made very small while keeping a constant pumped
charge per cycle, for non-adiabatic pumping through a system with well separated resonances, the frequency scaling of the pumped current is
typically found to be quadratic
\cite{Vavilov2001,FoaTorres2005,Agarwal2007}. In the main frame of
Fig. 2, we show the frequency scaling of the maximum pumped current
(discriminated by subband) in a broad range of frequencies going
from tens of GHz to THz. 
Although, no simple behavior is expected for the maximum current due to a complicated interplay between finite size effects
and driving parameters, Fig. 2 shows a dependence on frequency for $q=7,9$ which is for lower frequencies definitely better than the quadratic one observed for $q=8$.  
Notwithstanding, all the curves show a decrease for high enough frequency. This is shown only for $q=9$ in the plot where the current reveals a maximum value. This behavior can be understood as an interplay between
the enhancement of inelastic processes for low frequency as the vHS becomes prominent
and the opposite tendency produced when the frequency is large enough such that the field becomes ineffective in exciting photons.
The same pumping behaviour was verified for
armchair graphene nanoribbons, the only difference being the precise
position of the vHS and the number of available channels.

\textit{Conclusions}. Most of the studies on quantum charge pumping
focus on the behavior of a two-parameter pump close to isolated resonances. Here we focused
on a different scenario where non-adiabatic effect dominate: single-parameter pumping in a system with very good contacts. Our results are
illustrated for two cases of much experimental relevance: carbon
nanotubes and graphene nanoribbons. Besides offering currents that
are higher than those driven close to the charge neutrality point,
pumping phenomena held close to a van Hove singularity renders an improved scaling 
with frequency. On the other hand, the pumped current close to the charge neutrality point may also be substantially enhanced by tuning the driving parameters and system size. 
%instead of the typical quadratic dependency.

This work was supported by the Alexander
von Humboldt Foundation, the European Union project
CARDEQ under contract No. IST-021285-2, SeCyT-UNC, CONICET (Argentina) and ANPCyT. GC acknowledges support from the South Korean Ministry of Education, Science, and Technology Program, Project WCU ITCE No. R31-2008-000-10100-0. Computing time provided by the ZIH at the Dresden University of Technology is also acknowledged. 
%\bibliographystyle{prsty}
%\bibliography{bibAC}

\end{document}